\documentstyle[thmsa,a4,sw20lart]{article}
\newlength{\defaultparindent}
\newenvironment{
1}{}{}
\newenvironment{
1.22 1}{}{}
\newenvironment{Oleg 1}{}{}
\newenvironment{ Page Number}{}{}
\newenvironment{Default Paragraph Font}{}{}
\input{tcilatex}
\begin{document}

\author{A. M. Glukhov\thanks{%
B. Verkin Institute for Low Temperature Physics and Engineering of the
National Academy of Sciences of Ukraine, 47 Lenin Ave., Kharkov 310164,
Ukraine}, A. S. Pokhila\footnotemark[1]  , I. M. Dmitrenko\footnotemark[1] 
, \and A. E. Kolinko\footnotemark[1]  and A. P. Panchekha\thanks{%
Physics Department, Queens College, 65-30 Kissena Blvd, Flushing, NY 11367,
USA}}
\title{Superconducting Quantum Interference in Fractal Percolation Films. Problem
of 1/f Noise }
\date{}
\maketitle

\begin{abstract}
An oscillatory magnetic field dependence of the DC voltage is observed when
a low-frequency current flows through superconducting Sn-Ge thin-film
composites near the percolation threshold. The paper also studies the
experimental realisations of temporal voltage fluctuations in these films.
Both the structure of the voltage oscillations against the magnetic field
and the time series of the electric ''noise'' possess a fractal pattern.
With the help of the fractal analysis procedure, the fluctuations observed
have been shown to be neither a noise with a large number of degrees of
freedom, nor the realisations of a well defined dynamic system. On the
contrary the model of voltage oscillations induced by the weak fluctuations
of a magnetic field of arbitrary nature gives the most appropriate
description of the phenomenon observed. The imaging function of such a
transformation possesses a fractal nature, thus leading to power-law spectra
of voltage fluctuations even for the simplest types of magnetic fluctuations
including the monochromatic ones. Thus, the paper suggests a new universal
mechanism of a ''1/f noise'' origin. It consists in a passive transformation
of any natural fluctuations with a fractal-type transformation function.
\end{abstract}

\section{Introduction}

Studies of the object critical behavior near the superconducting phase
transition or a metal-dielectric transition are related to the fundamental
problems of describing systems characterized by a scale or time invariance,
or so-called fractal (self-similar) systems \cite{Mand1}.

The 1/f noise, or the ``flicker noise``, remains a very important and
enigmatic problem which is far from its solution. The fact that a noise of
the 1/f type is found practically everywhere (in physical, biological and
even in social systems) stimulates the search of some general approaches to
this problem. The language of fractals is just such an approach, and a
number of theoretical works on the ``flicker noise`` have been written in
this language. In particular, there appeared a theory of self-organized
criticality \cite{PerBak} in which it was shown that a dissipative dynamic
system evolves naturally to a critical state where characteristic time and
length scales are absent. Such states possess a scale-invariant (fractal)
structure and undergo fluctuations with a spectrum of the 1/f type.

There are not enough experimental works carried out in this field, because
of the difficulties of obtaining model fractal systems and of observing
sufficiently distinct fluctuations of physical quantities in them. In this
connection, quantum interference effects in macroscopically inhomogeneous
media look rather promising.

An AC current passing through systems with weak bonds in an applied magnetic
field $H$ at a temperature below the superconducting transition temperature
is known to induce a DC voltage $V_{DC}$ oscillating against a variation of
the field $H$ \cite{Ouboter}. Such oscillatory dependences $V_{DC}(H)$ were
observed by Yurchenko et al. \cite{Yurch} in Nb powders, and in the Clarke
drop-type interferometers \cite{Clarke}. The dependences were attributed to
the quantum interference effects in asymmetrical contours with a weak
coupling, i.e. the magnetic flux quantization induces the critical current
oscillations and the respective voltage oscillations. According to De Waele
and De Bruyn Ouboter \cite{Ouboter}, when an AC current passes through a
system of two superconductors weakly connected by an asymmetric double-point
contact with an amplitude larger than the critical current of the junction,
then a DC voltage is observed even in the absence of the DC current . This
rectification process takes place due to the fact that the absolute values
of the voltage oscillations are shifted with respect to each other for
different signs of a DC current. The resultant DC voltage depends
periodically on the applied magnetic field with a period $\Delta B_{\bot
}=h/2eO$ ($O$ being the area enclosed between the two contacts) and it is
antisymmetric under field reversal. A superconducting granulated film at the
percolation threshold can be obviously considered as a random set of
asymmetric Josephson contours. In this connection, as observed by Gerber and
Deutscher \cite{Gerber}, DC voltage oscillations in granulated Pb and Al
percolation films could be explained by the superconducting quantum
interference.

Thus, the DC voltage oscillations in percolation films have been associated
with the topological structure of quantizing contours; however, nobody has
ever tried to relate them to the sample fractal structure. Thus, the idea
has originated to investigate correlations between the structure of quantum
voltage oscillations in a magnetic field and topological properties of
thin-film superconducting composites in the vicinity of the percolation
threshold, within the framework of the fractal approach, in order to
elucidate the mechanism of the ``1/f noise`` origin in self-similar objects.

Thus, two curious phenomena, observed in granulated percolation thin film
Sn-Ge systems, form the object of our investigations:

1. Fractal character of voltage $V_{DC}$ induced by an AC current passing
through system as a function of the external magnetic field $H$ - $V_{DC}(H)$%
.

2. Time fluctuations of the voltage at a constant value of the magnetic
field $H$.

\section{Experiment}

Granulated films for the Sn - amorphous Ge system with monotonically varying
structural characteristics are made by vacuum condensation of Sn on a long ($%
60mm$) substrate along which a temperature gradient is created. The
temperature at the midpoint of the substrate ($80^{\circ }$C) corresponds to
the temperature of the Sn condensation mechanism change. Sn is deposited on
the previously prepared Ge layer $50nm$ thick. The effective thickness of
the Sn layer is $60nm$. The metallic condensate is covered with amorphous Ge
from the top too.

On using this technique granulated films are obtained. The films have a
varying (along the substrate) structure: near the ``cool'' end they have a
labyrinthine structure with low resistance ($1\Omega $ per square), while at
the ``hot'' end they have an island structure with high resistance ($%
80k\Omega $ per square). This structural change results in the
metal-dielectric transition within one series, consisting of $30$ samples,
the percolation threshold being near the substrate midpoint. The technique
of obtaining granulated films and their resistive properties on the metal
and dielectric sides of the metal-nonmetal transition are described in
detail in Refs \cite{Glukh1, Glazman}. For the present investigations, we
chose the films near the percolation threshold with a characteristic
structure depicted in Fig.\ref{fig1}.

\FRAME{ftbpFU}{3.1332in}{3.2629in}{0pt}{\Qcb{Electron micrograph of Sn-Ge
sample near the percolation threshold. Light regions correspond to metal.}}{%
\Qlb{fig1}}{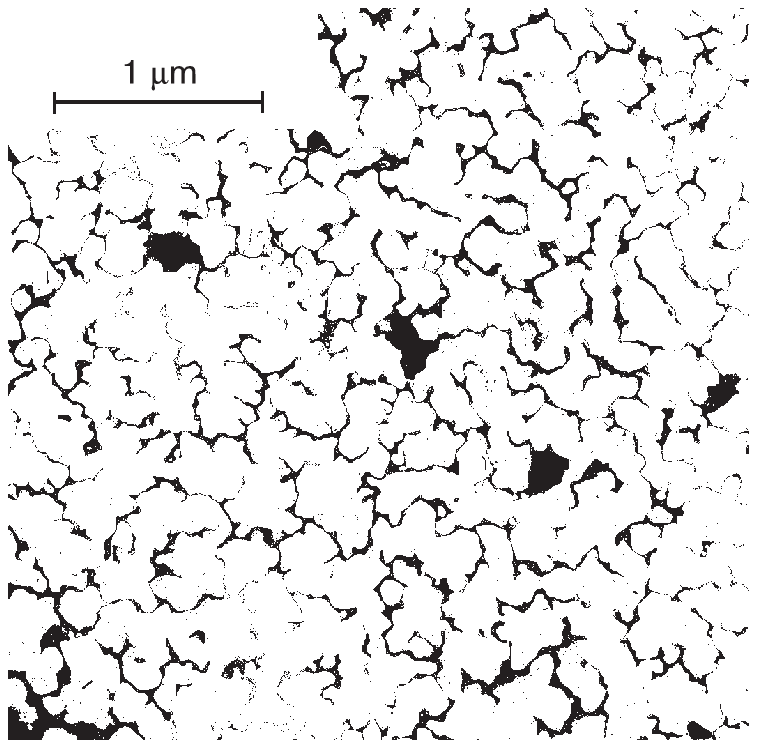}{\special{language "Scientific Word";type
"GRAPHIC";maintain-aspect-ratio TRUE;display "USEDEF";valid_file "F";width
3.1332in;height 3.2629in;depth 0pt;original-width 215.8125pt;original-height
224.8125pt;cropleft "0";croptop "1";cropright "1";cropbottom "0";filename
'FIG1.EPS';file-properties "XNPEU";}}

To achieve our aim we carried out experiments of two types:

(a) registration of rectified voltage as a function of an external magnetic
field while passing AC through the sample, $V_{DC}(H)$;

(b) recording the low-frequency part of ``noises'' in the sample while
passing DC through it in the absence of an external magnetic field, $V(t)$.

The electric measurements were carried out according to the standard
four-probe technique. The DC was measured to an accuracy of $0.01\%$ in the
range of $0.01$ - $2mA$ for various samples, and AC of $1$ - $200kHz$
frequency, $0.2$ - $2mA$ effective value and $0.1\%$ stability was provided
by a generator of sinusoidal signals through a decoupling transformer. The
voltage in the both experiments was registered by a digital voltmeter with
the measuring frequency amounting to $12$ measurements per second and the
accuracy of $0.1\mu V$. The usual length of recording a noise dependence was 
$8192$ points. Input and processing of experimental data were made with the
help of the original computer algorithms.

Samples were immersed in liquid He inside a superconducting solenoid. The
measuring cell construction enabled one to vary the sample orientation
relative to the magnetic field direction. To avoid possible hysteresis
phenomena, the dependence $V_{DC}(H)$ was recorded with various directions
of the current through the solenoid. The scanning step in the magnetic field
varied, and did not exceed $10^{-6}T$. In the operating temperature range $%
2.5$ - $4.2K$ the temperature stabilization was not worse than $10^{-3}K$
with the superconducting transition width of $0.2$ - $1.0K$. To avoid a
possible effect of temperature fluctuations, temperature values were
simultaneously recorded with an accuracy of $2\times 10^{-4}K$, but no
noticeable drifts or fluctuations were found.

\section{Results and discussion}

\subsection{Fractal analysis of the structure of superconducting quantum
voltage oscillations}

As it was described in our brief report \cite{Glukh2}, the samples under
investigation were characterized by highly extended superconducting
transitions. At the temperatures below the midpoint of the resistive
transition, clearly manifested oscillations of the DC voltage $V_{DC}$ were
observed against the magnetic field $H$ applied, when an AC current passed
through the sample. The amplitude and frequency of the current did not
affect the form of the $V_{DC}(H)$ dependence significantly. The results
could be easily reproduced. Fig.\ref{fig2} shows the $V_{DC}(H)$ dependence
for various orientations of the film relative to the magnetic field. The
scale of the oscillatory structure in the field is inversely proportional to
the cosine of the angle between the applied magnetic field and the normal to
the sample plane. The emergence of the normal magnetic field component alone
as well as the antisymmetry of the oscillatory structure relative to $H=0$
indicate the quantum-interference origin of $V_{DC}(H)$.

\FRAME{ftbpFU}{3.0753in}{2.3808in}{0pt}{\Qcb{ Oscillatory structure of
voltage across a Sn-Ge sample for various angles $\theta $ between the
applied magnetic field and the normal to the surface. $T=3.354K$, $%
I_{AC}=0.2mA$, $f=100kHz$.}}{\Qlb{fig2}}{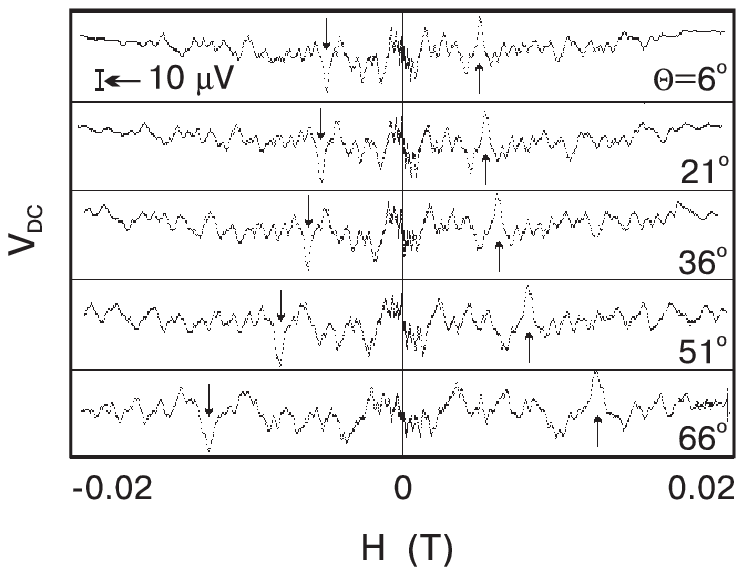}{\special{language
"Scientific Word";type "GRAPHIC";maintain-aspect-ratio TRUE;display
"USEDEF";valid_file "F";width 3.0753in;height 2.3808in;depth
0pt;original-width 211.8125pt;original-height 163.625pt;cropleft "0";croptop
"1";cropright "1";cropbottom "0";filename'FIG2.EPS';file-properties "XNPEU";}%
}

\FRAME{ftbpFU}{3.1038in}{2.3808in}{0pt}{\Qcb{Magnetic field dependences of $%
V_{DC}$ at various temperatures.}}{\Qlb{fig3}}{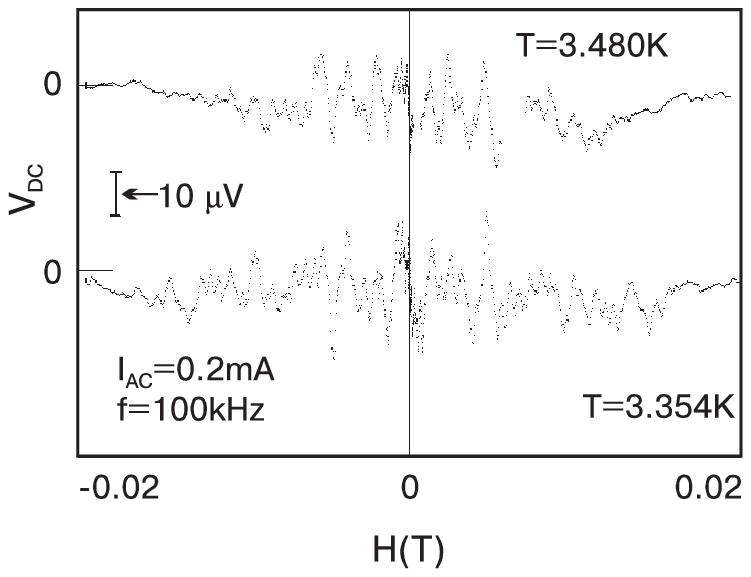}{\special{language
"Scientific Word";type "GRAPHIC";maintain-aspect-ratio TRUE;display
"USEDEF";valid_file "F";width 3.1038in;height 2.3808in;depth
0pt;original-width 213.8125pt;original-height 163.625pt;cropleft "0";croptop
"1";cropright "1";cropbottom "0";filename'FIG3.EPS';file-properties "XNPEU";}%
}

A decrease in temperature leads to new oscillations in the case of strong
fields (Fig.\ref{fig3}), and the mean square deviation $S$ of the rectified
voltage which characterizes the average amplitude of oscillations, has a
maximum near $2.6K$. Such a behavior of $V_{DC}(H)$ and $S$ is associated
with an increase in the number of quantizing contours and with an increase
in their critical fields with decreasing temperature. A further decrease in
temperature causes a decrease in $S$ due to closure of the percolation net
and a decrease in the number of weak bonds.

In the system under investigation, quantizing contours belong to the
skeleton of the metal percolation cluster. A percolation system near the
percolation threshold can be regarded as fractal, self-similar, on the
scales smaller than the correlation length $\xi $, and as homogeneous on
larger scales \cite{Stauffer, Kapit}. In the homogeneous case, the
dependence of the infinite cluster mass contained in a $a\times a$ square is
given on this scale $a$ by the expression $M_{\symbol{126}}a^{d}$, where $d$
is the conventional topological dimension. When $a<<\xi $, an anomalous
behavior of mass is observed in the self-similar mode, $M_{\symbol{126}%
}a^{D} $. $D$ is called the fractal dimension. Density, in its turn, behaves
as

\begin{equation}
\rho =\frac{M}{a^{d}}=\QATOPD\{ . {a^{D-d}\text{ for }a<\xi ,}{const\text{
for }a>\xi ,}  \label{1}
\end{equation}
i.e. on scales larger than $\xi $ the density becomes constant. The theory
predicts the relation $D=d-\beta /\nu $. Here $\beta $ and $\nu $ are
critical exponents of the infinite cluster density and the correlation
length, respectively. In the 2D case $\beta =0.14$, $\nu =1.33$, so that $%
D=1.896$.

To analyze the fractal dimensions of real objects, we processed electronic
microscope pictures of the investigated Sn-Ge samples on a computer. The
procedure is the following. First, we choose an initial point in the
computer picture of the percolation cluster (Fig.\ref{fig1}). Next, the
masses $M$ (i.e. quantities of white points) of the infinite cluster are
counted; the counted points must lie in the square centered at the chosen
initial point and $2a$ large, from $0$ to the picture boundary. This
operation is repeated several times for different position of the initial
point on the infinite cluster. After averaging over all the initial points,
the quantity $M(a)$ together with the density values $\rho (a)=M(a)/a^{2}$
is presented in the double logarithmic coordinates (Fig.\ref{fig4}). The
radius, along which the fractal behavior is replaced by the homogeneous one,
corresponds to the percolation length of the correlation, and equals about
60 points for the sample considered. By the least-squares method, we find
the slope of the dependence $\ln (M)$ on $\ln (a)$ in the fractal and
homogeneous modes, which correspond to the fractal $D=1.88$ and conventional 
$d=2.005$ dimensions. One can see that the fractal dimension value agrees
well with the theory.

\FRAME{ftbpFU}{3.1185in}{2.7856in}{0pt}{\Qcb{The dependences of the infinite
cluster mass $M$ and density $\rho $ on linear size of sample under
investigation (see Fig.\ref{fig1}).}}{\Qlb{fig4}}{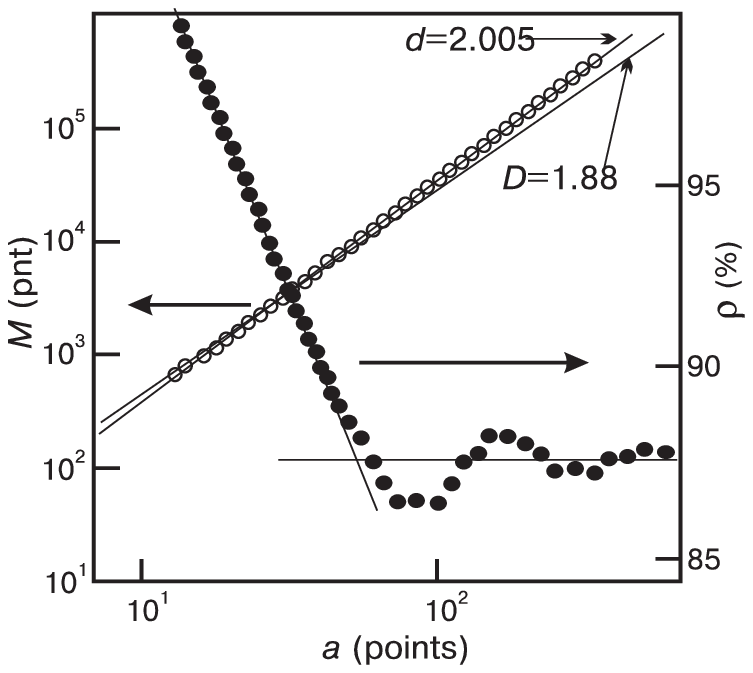}{%
\special{language "Scientific Word";type "GRAPHIC";maintain-aspect-ratio
TRUE;display "USEDEF";valid_file "F";width 3.1185in;height 2.7856in;depth
0pt;original-width 214.8125pt;original-height 191.6875pt;cropleft
"0";croptop "1";cropright "1";cropbottom
"0";filename'FIG4.EPS';file-properties "XNPEU";}}

In view of the fact that the scale of oscillations in a magnetic field under
the conditions of quantum interference is unambiguously connected with the
size of quantizing contours, we can assume that voltage oscillations
occurring at a percolation fractal cluster have a fractal structure too. The
fractal dimensions of the plots of the $V_{DC}(H)$ dependence can be
calculated formally by covering them with cells of width $bh$ along the
magnetic field axis and of length $bv$ along the voltage axis so that the
minimum size of a cell is $h\times v$. Then the fractal dimension $D_{c}$
over the covering is determined from the following dependence of the number $%
N_{c}$ of cells required for covering the curve on the scale $b$ of the
cells: $N_{c}(b;v,h)_{\symbol{126}}b^{-D_{c}}$ \cite{Mand2}. The minimum
cell width is $h=1$, and for the height $v$ we can take a value of the order
of resolution of the measuring voltmeter. We calculated the cell dimensions
of the $V_{DC}(H)$ dependence at all temperatures. The dependence of the
number $N_{c}$ of cells covering the $V_{DC}(H)$ curve on the cell scale $b$
for a certain value of temperature is presented in Fig.\ref{fig5} in
logarithmic coordinates. The fractal dimensions $D_{c}$ obtained by
approximating the experimental data were found to be $1.6$ - $1.7$. They
revealed no systematic dependence on temperature and practically coincided
with the theoretical fractal dimension of the skeleton of a two-dimensional
percolation cluster ($1.62$) \cite{Stauffer}.

\FRAME{ftbpFU}{3.1332in}{2.2511in}{0pt}{\Qcb{Determination of cellular
fractal dimensions of the $V_{DC}(H)$ curves. The dependence of the number $%
N_{c}$ of cells covering the $V_{DC}(H)$ on the cell scale $b$.}}{\Qlb{fig5}%
}{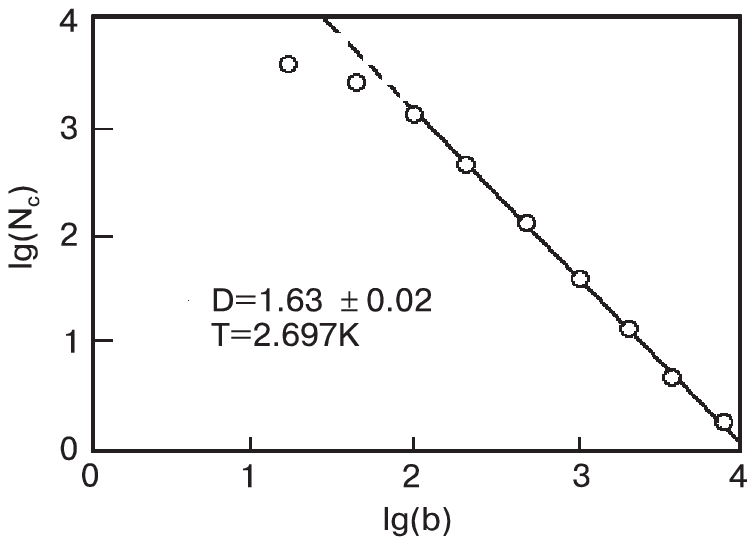}{\special{language "Scientific Word";type
"GRAPHIC";maintain-aspect-ratio TRUE;display "USEDEF";valid_file "F";width
3.1332in;height 2.2511in;depth 0pt;original-width 215.8125pt;original-height
154.5625pt;cropleft "0";croptop "1";cropright "1";cropbottom "0";filename
'FIG5.EPS';file-properties "XNPEU";}}

That the $V_{DC}$ oscillatory structure is actually fractal is additionally
confirmed by the computations using the rescaled range method, or $R/S$
analysis, also called the Hurst method \cite{Hurst}. Its essence consists in
studying the normalized (by the standard deviation $S$) difference $R$
between the maximum and minimum accumulated deviations of the random value
from its mean. As Hurst found out, the observed dependence of the normalized
range $R/S$ on the excerpt length $L$ describing this range, for various
natural processes is well fitted by a power law $R/S_{\symbol{126}}L^{K_{H}}$
with the exponent $K_{H}$ close to $0.7$. At the same time, if the temporal
series are related to random processes with independent values and a finite
variance, then $K_{H}=0.5$. As shown by Mandelbrot \cite{Mand2} the exponent 
$K_{H}$ is related to the fractal dimension $D_{B}$ in covering the
accumulated deviation by the formula

\begin{equation}
D_{B}=2-K_{H}  \label{2}
\end{equation}
for self-affine curves.

\FRAME{ftbpFU}{1.9346in}{1.5506in}{0pt}{\Qcb{The dependence of the number $%
N_{c}$. of cells covering the accumulated deviation of the $V_{DC}(H)$ curve
on the cell scale $b$ and diagram of $\log (R/S)$ versus $\log (L)$.}}{\Qlb{%
fig6}}{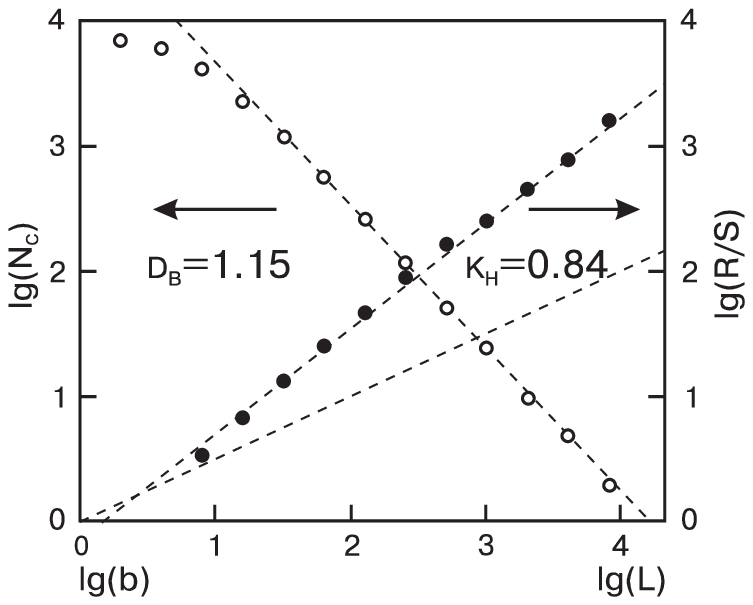}{\special{language "Scientific Word";type
"GRAPHIC";maintain-aspect-ratio TRUE;display "USEDEF";valid_file "F";width
1.9346in;height 1.5506in;depth 0pt;original-width 214.8125pt;original-height
171.625pt;cropleft "0";croptop "1";cropright "1";cropbottom "0";filename
'FIG6.EPS';file-properties "XNPEU";}}

Fig.\ref{fig6} shows the result of the $R/S$ analysis program applied to an
example of voltage fluctuations on a Sn-Ge sample in the vicinity of the
superconducting transition. The $R/S$ value is first calculated for the
entire excerpt, then the latter is halved, and two independent $R/S$ values
are obtained that characterize the half-excerpts separately. Then the number 
$L$ of points is halved again and again, until it becomes less than $8$; at
each step the number of independent regions is doubled. The results that
correspond to the same value of $L$ are averaged and the $R/S$ values are
plotted on the double-logarithmic scale as a function of the number of
points $L$. The exponent $K_{H}$ is obtained by the least-squares method. In
the figure $K_{H}=0.84\pm 0.01$. The broken line shows the dependence at $%
K_{H}=0.5$, i.e. ``white'' noise. One can see in Fig.\ref{fig6} that $%
D_{B}=1.15\pm 0.01$, which agrees with expression \ref{2}. All these data
confirm that the oscillatory structure of $V_{DC}(H)$ is actually fractal.

The observed oscillatory structure is a superposition of quantum
oscillations of voltage occurring in Josephson contours of various size up
to the geometrical size of the sample. The fractal (i.e., self-similar)
behavior of $V_{DC}(H)$ on all the scales in the magnetic field was also
confirmed by the Fourier analysis of the $V_{DC}(H)$ dependence. The $%
V_{DC}(H)$ oscillation intensity spectrum (presented in Fig.\ref{fig7})
indicates the absence of clearly manifested peaks, viz., preferred magnetic
fields and, hence, quantizing contours.

\FRAME{ftbpFU}{3.109in}{2.2701in}{0pt}{\Qcb{A typical spectrum of voltage
oscillation power $P$ in magnetic field. The area $a^{2}$ of the
quantization contour plays the role of frequency ($1/H$ in our case).}}{\Qlb{%
fig7}}{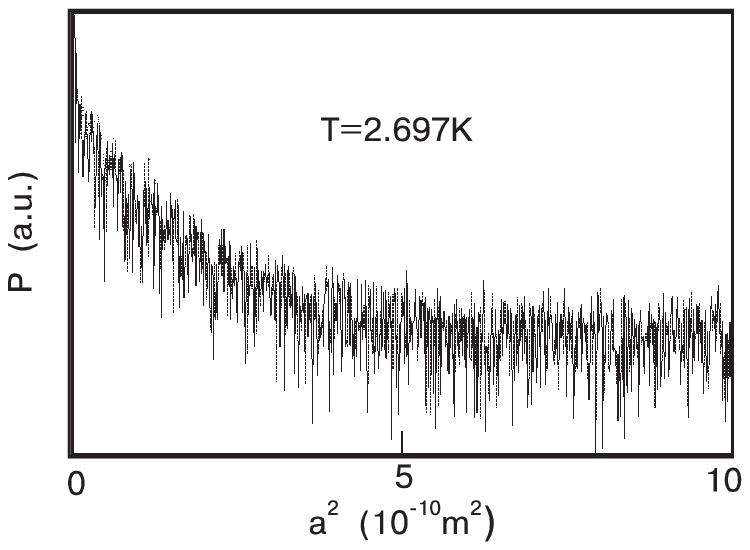}{\special{language "Scientific Word";type
"GRAPHIC";maintain-aspect-ratio TRUE;display "USEDEF";valid_file "F";width
3.109in;height 2.2701in;depth 0pt;original-width 213.8125pt;original-height
155.5625pt;cropleft "0";croptop "1";cropright "1";cropbottom "0";filename
'FIG7.EPS';file-properties "XNPEU";}}

\subsection{Fractal analysis of voltage time series (electric ``noise'')}

The electric noise and, in particular, the ``1/f noise'', observed in
percolation systems seems to be determined by the percolation fractal
geometry of the systems. Hence, it is reasonable to use the fractal language
to the description of random time series and to the search of promising
mathematical model of the mechanism of the ``1/f noise'' origination.

First, we will describe the procedure we used to interpret experimental data.

\FRAME{ftbpFU}{1.497in}{3.6979in}{0pt}{\Qcb{Time series of voltage
fluctuations at $T=3.607K$ (a), its energy spectrum (b) and the
autocorrelated function (c) ($t_{0}=85ms$ - counting period).}}{\Qlb{fig8}}{%
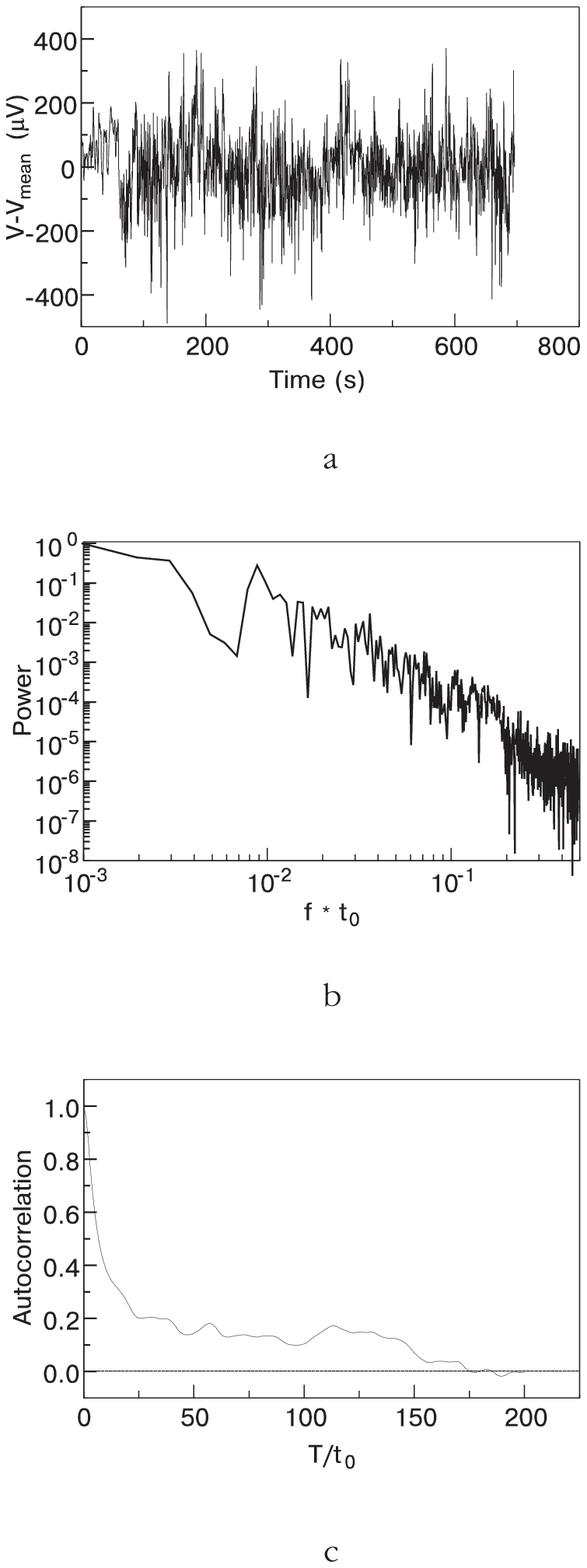}{\special{language "Scientific Word";type
"GRAPHIC";maintain-aspect-ratio TRUE;display "USEDEF";valid_file "F";width
1.497in;height 3.6979in;depth 0pt;original-width 232.875pt;original-height
580.1875pt;cropleft "0";croptop "1";cropright "1";cropbottom "0";filename
'FIG8.EPS';file-properties "XNPEU";}}

Fig.\ref{fig8} shows a typical time series (Fig.\ref{fig8}a), its energy
spectrum (Fig.\ref{fig8}b) and the autocorrelated function (Fig.\ref{fig8}%
c). One can see that the fluctuations under study have a sizable correlation
on large scales, which narrows the range of feasible mathematical models.
Actually, there remain with two ways that permit constructive testing:

- the model of the colored noise with a power-law spectrum and random phases;

- the model of a randomized dynamic system with few degrees of freedom (at
least, for the high-frequency part of the spectrum).

Both the hypotheses can be verified by using the same procedure of fractal
analysis \cite{Akhrom}, which is able, as a minimum, to reject one or
another model. Let us consider this opportunity in more detail.

In this case under fractal analysis we mean the following sequence of steps:

- projecting the experimental scalar realization into the space of dimension 
$m=1,2,3...$ (imbedding dimension) by using the Takens procedure \cite
{Akhrom, Takens, Litvin};

- calculating the correlation integral $C(r,m)$ by using the Grassberger
procedure \cite{Grassb} for each projection;

- finding the scaling regions and determining the correlated dimension $%
D_{2}(m)$ for each projection;

- investigating the dependence of the correlated dimension $D_{2}$ on the
imbedding dimension and other parameters.

As is known, in the limit of the infinite realization length $L$ the
correlated dimension is determined by the relation

\begin{equation}
D_{2}(m)=\stackunder{L\rightarrow \infty }{\lim }\lim_{r\rightarrow 0}\frac{%
d(\log C(r,m))}{d(\log r)}  \label{3}
\end{equation}

Practically, for a finite realization one determines regions of linear
dependence (scaling regions) $\log (C)$ versus $\log (r)$, whose slope
yields the estimate of the value $D_{2}$:

\begin{equation}
\log (C(r,m))=D_{2}(m)\log (r)+A\text{ \ }for\text{ }r\text{ }in\text{ }[%
r_{1}...r_{2}]  \label{4}
\end{equation}

It was shown \cite{Takens} that one should expect and account for scaling
regions not smaller than one decimal order in $C(r_{2})/C(r_{1})$, at least,
for the imbedding dimension greater than $5$. Systematic and random errors
in estimating $D_{2}$ were evaluated in \cite{Takens} as well.

It is most essential for the present work that in realizations of
finite-dimensional dynamic systems the $D_{2}$ dependence on $m$ has a
typical shape of a plateau (Fig.\ref{fig9}) when $m>D$, and if the plateau
is absent, it enables one to reject the conjecture of a dynamic origin of a
realization.

\FRAME{ftbpFU}{3.1038in}{2.1344in}{0pt}{\Qcb{$D_{2}$ dependence on $m$ at $%
Lf_{0}=200$ for typical realizations registered at various temperatures.}}{%
\Qlb{fig9}}{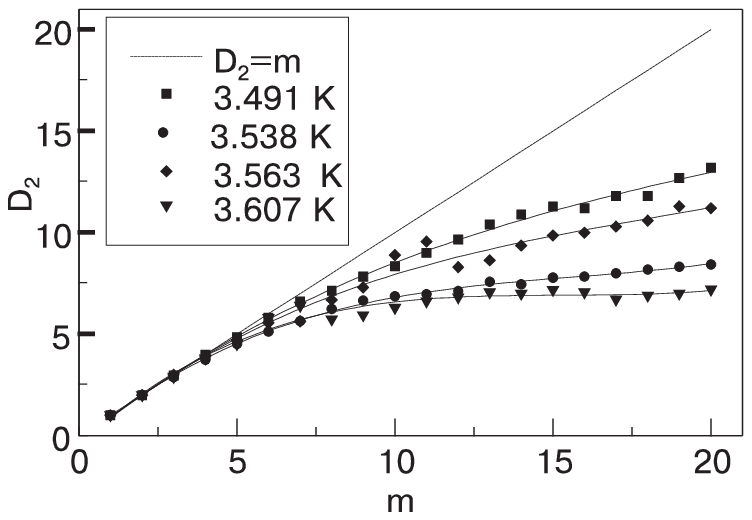}{\special{language "Scientific Word";type
"GRAPHIC";maintain-aspect-ratio TRUE;display "USEDEF";valid_file "F";width
3.1038in;height 2.1344in;depth 0pt;original-width 213.8125pt;original-height
146.5625pt;cropleft "0";croptop "1";cropright "1";cropbottom "0";filename
'FIG9.EPS';file-properties "XNPEU";}}

On the other hand, applications of this procedure to a realization of the
colored noise with the power spectrum and random phases were studied in Refs.%
\cite{Osborn, Theiler2}. It was shown \cite{Theiler2} that the results
depended substantially on the ratio of the realization length $L$ and the
lowest frequency $f_{0}$ of the realization spectrum. Namely, at $Lf_{0}<100$
there is a fictitious plateau on the $D_{2}(m)$ dependence; the plateau
disappears in the opposite case where $Lf_{0}>100$ (the threshold value
varies, depending on the exponent of the power spectrum, from one to several
hundreds). The physical sense of this phenomenon is related to the
manifestation of the signal non-stationarity at $Lf_{0}<100$. Thus, the
colored noise conjecture can also be, as a minimum, rejected, if the $D_{2}$
dependence on $Lf_{0}$ behaves not in the described fashion, namely when the
plateau at $Lf_{0}>100$ disappears. This analysis is similar to checking the
``null hypothesis'', which is mentioned in the literature \cite{Theiler2}.

In our case, the realization energy spectrum is obviously a power-law one,
all the way down to zero frequency. Varying the parameter $Lf_{0}$ can be
done by filtering realizations with a symmetric non-recursive filter which
gives way only to the high-frequency spectrum part $f>f_{0}$. So in this
paper we investigate filtered realizations of the initial set by applying
the fractal analysis. In each realization we analyze the dependence of the
correlated dimension $D_{2}$ on

- the imbedding dimension $m$ at a fixed parameter $Lf_{0}$,

- the parameter $Lf_{0}$ at a fixed imbedding dimension $m$.

Here we shall describe and analyze some results of the above-described
procedure to the experimentally obtained realizations of voltage
fluctuations registered at various temperatures near and below the mean
point of the superconducting transition.

In all realizations the imbedding dimension $m$ varied within the range $%
1..20$, while the non-stationarity parameter $Lf_{0}$ took on values $1$, $%
40 $, $100$, $200$, $400$ at the realization length of $8100$ points. The
separation between elements of one vector in the Takens procedure was taken
equal to the position of the first zero of the autocorrelated function,
while in the Grassberger procedure vectors were excluded if the time
separation between them was less than this value; this was done to exclude
the so-called ``shoulder'' effect \cite{Theiler1}. Fig.\ref{fig9} shows a
plot of the $D_{2}$ dependence on $m$ at $Lf_{0}=200$ for typical
realizations registered at various temperatures. As seen in the figure, the
dependence has a pronounced plateau at rather moderate values of $D_{2}$.
This confirms the reliable registration of the non-random nature of the
realization under study. In other words, the conjecture of the dynamic
nature of the studied realizations cannot, as a minimum, be rejected.

An additional confirmation of the latter statement is provided in Fig.\ref
{fig10}, which demonstrates the change in the $D_{2}(m)$ dependence on
varying the parameter $Lf_{0}$. (Here we have the same realization $T=3.607K$
which was presented in Fig.\ref{fig8}a). One can see that at $Lf_{0}>100$
the plateau displays no tendency to disappearing. Thus, the result of the
analysis enables us to reject the hypothesis of colored noise with a
power-law spectrum and random phases.

\FRAME{ftbpFU}{3.1185in}{2.0626in}{0pt}{\Qcb{Dependence of the correlated
dimension $D_{2}$ on the imbedding dimension $m$ at $T=3.607K$ on varying
the parameter $Lf_{0}$.}}{\Qlb{fig10}}{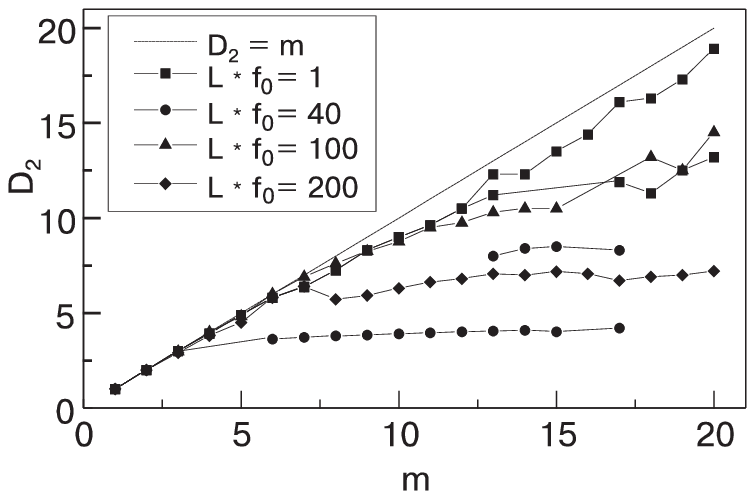}{\special{language
"Scientific Word";type "GRAPHIC";maintain-aspect-ratio TRUE;display
"USEDEF";valid_file "F";width 3.1185in;height 2.0626in;depth
0pt;original-width 214.8125pt;original-height 141.5pt;cropleft "0";croptop
"1";cropright "1";cropbottom "0";filename 'FIG10.EPS';file-properties
"XNPEU";}}

The only circumstance that remains unexplained is the lowering of the
plateau on the $D_{2}(m)$ dependence when the parameter $Lf_{0}$ increases
beyond the threshold value $Lf_{0}=100$.

In principle, in the spirit of Ref.\cite{Gapon}, one could attempt, to
interpret such a behavior as a mixture of several dynamic systems with
different characteristic frequencies, whose relative weights vary with
changing the threshold frequency of the filter $f_{0}$. We believe, however,
that in this case such an interpretation would be too arbitrary. That is why
we have attempted to interrelate spatial and temporal dependences typical
just for this case. We suggest a simple dynamic model which accounts for our
experimental data without loss of any details.

\section{A model of ``1/f noise'' generation in a percolation system.}

In the preceding sections, we have presented results of investigating the
fractal nature of the dependences $V_{DC}(H)$, and analyzed the results of
applying the fractal analysis to experimental time realizations of voltage
fluctuations in superconducting percolation films near the superconducting
transition temperature. In particular, we have presented the arguments for
inadequacy of the colored random noise and for the adequacy of the
deterministic dynamic model in a stochastic mode. However, the correlated
dimension dependence on the filtration parameter $Lf_{0}$ has remained
unclear.

In the percolation system under study, another, better studied, phenomenon
is observed which has a fractal nature and can have a relation to the
investigated fluctuations. The fractal nature of the dependence $V_{DC}(H)$
was investigated in detail above via the calculation of its cell dimension
and the $R/S$ analysis, as well as by direct scale transformations. Since
the internal cell dimension of the plot of $V_{DC}(H)$ tends to $1.6$-$1.7$,
while that of the accumulated signal tends to the stationary value near $%
1.15 $, we can conclude that the plot of $V_{DC}(H)$ is close to the
pointwise approximation of the fractal curve derivative.

Suppose that in the sample there are oscillations of a magnetic field, of
internal or external origin (Fig.\ref{fig11}c). Then the passive detection
of these oscillations with the fractal transition function of detector $%
V_{DC}(H)$ (Fig.\ref{fig11}a) will contribute to the voltage across the
sample (Fig.\ref{fig11}b). Here two important peculiarities will be
observed. First, since the fractal curve derivative can have arbitrarily
large slope angles (to within the accuracy owing to its finite approximation
determined by the geometric dimensions of the sample), any signal, however
weak, can be amplified to the maximum value while detecting it. Second, the
multi-scale pattern of the fractal curve will impose a power-law spectrum on
very simple oscillations detected.

\FRAME{ftbpFU}{3.1185in}{2.2943in}{0pt}{\Qcb{Passive transformation of
magnetic field oscillations to electric ``1/f noise''.}}{\Qlb{fig11}}{%
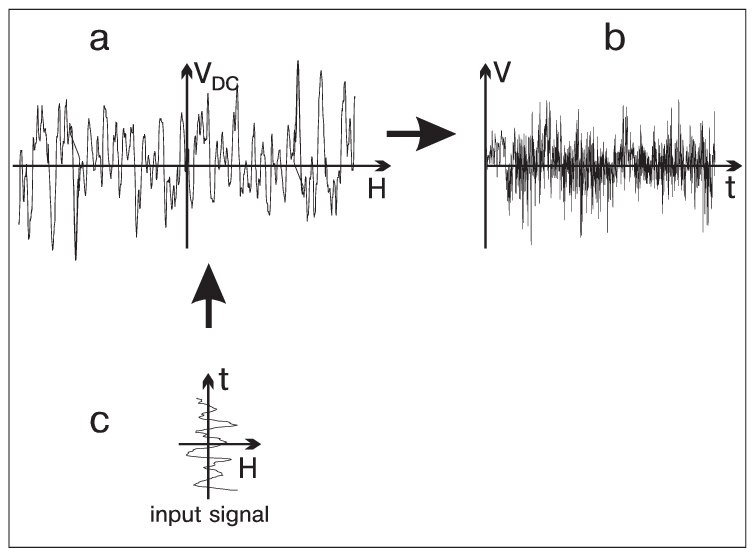}{\special{language "Scientific Word";type
"GRAPHIC";maintain-aspect-ratio TRUE;display "USEDEF";valid_file "F";width
3.1185in;height 2.2943in;depth 0pt;original-width 214.8125pt;original-height
157.5625pt;cropleft "0";croptop "1";cropright "1";cropbottom "0";filename
'FIG11.EPS';file-properties "XNPEU";}}

In order to verify this model, we repeated the fractal analysis procedure
for a realization simulated as follows. First of all we modeled the
transition function $V_{DC}(H)$, using the finite-difference derivative of
the Weierstrass function $W(x)$, obtained by summing a finite number $K$ of
periodic terms.

\begin{equation}
V_{sim}(H)=W(H+\Delta )-W(H)  \label{5}
\end{equation}

\begin{equation}
W(x,K)=\sum\limits_{i=1}^{K}2^{-\alpha i}\cos (2^{ix})  \label{6}
\end{equation}
Then the model realization for voltage time fluctuations $E(t)$ has the form

\begin{equation}
E(t)=V_{sim}(h(t))\text{,}  \label{7}
\end{equation}
where $h(t)$ is the model of magnetic field oscillations of any kind. For
better visualization we tried the simplest functions

\begin{equation}
h(t)=\sin (t)\text{,}  \label{8}
\end{equation}

\begin{equation}
h(t)=\sin (t)+\sin \left( \frac{\pi t}{4}\right) \text{.}  \label{9}
\end{equation}

Here the parameter $\alpha $ was selected so that both the energy spectrum
of the realization $E(t)$, and the $R/S$ parameter of the transition
function $V_{sim}(H)$, as well as the cell dimension of accumulated signal
would coincide with the values calculated for one of the experimental
realizations. Both coincidences were reached by varying only the parameter $%
\alpha $; here the dependence on $\Delta $ and number $K$ of terms of the
trigonometric series appeared insignificant. Practically, we used the
following values: $\alpha =1.5$, $\Delta =10^{-4}$, $K=10$. Below we present
the results of analysis of a model realization with an external signal in
the form of Eq.\ref{9}. Here the energy spectrum of the model realization
(see Fig.\ref{fig12}) has the form quite similar to the natural data.

\FRAME{ftbpFU}{3.1185in}{2.1638in}{0pt}{\Qcb{The energy spectrum of a model
realization.}}{\Qlb{fig12}}{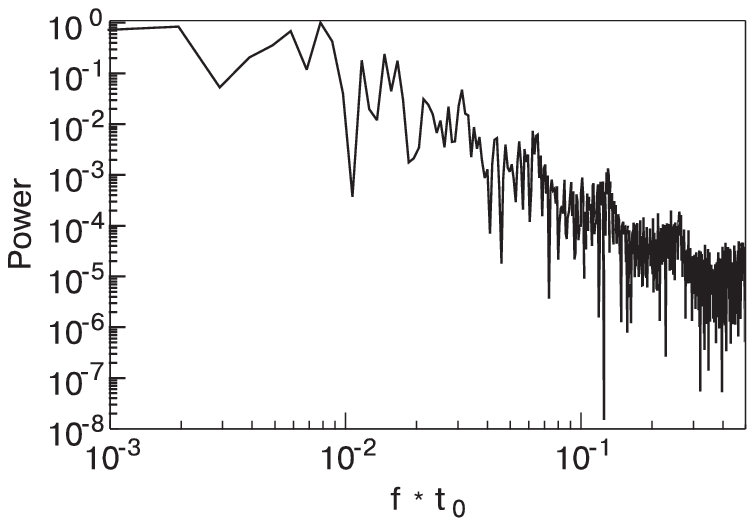}{\special{language "Scientific
Word";type "GRAPHIC";maintain-aspect-ratio TRUE;display "USEDEF";valid_file
"F";width 3.1185in;height 2.1638in;depth 0pt;original-width
214.8125pt;original-height 148.5625pt;cropleft "0";croptop "1";cropright
"1";cropbottom "0";filename 'FIG12.EPS';file-properties "XNPEU";}}

\FRAME{ftbpFU}{2.1248in}{3.6962in}{0pt}{\Qcb{$D_{2}$ dependence on $m$ for
the model realization at $Lf_{0}=100$ (a), $Lf_{0}=200$ (b).}}{\Qlb{fig13}}{%
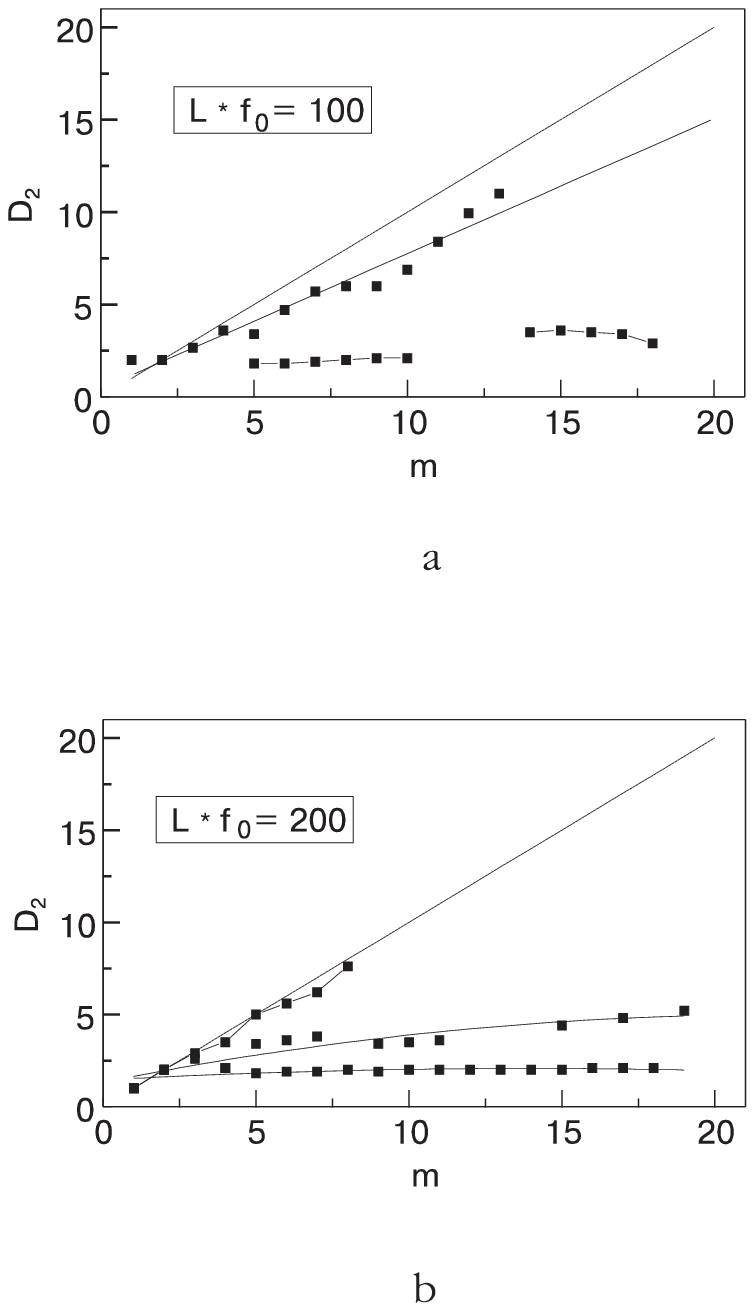}{\special{language "Scientific Word";type
"GRAPHIC";maintain-aspect-ratio TRUE;display "USEDEF";valid_file "F";width
2.1248in;height 3.6962in;depth 0pt;original-width 214.8125pt;original-height
375.375pt;cropleft "0";croptop "1";cropright "1";cropbottom "0";filename
'FIG13.EPS';file-properties "XNPEU";}}

In Figs.\ref{fig13}a and \ref{fig13}b the $D_{2}(m)$ dependence is shown for
various values of the parameter $Lf_{0}$. One can see that the dependence
has all the characteristic features of the experimental realizations shown
in Fig.\ref{fig10}. Namely, there are well-shaped plateaux on the
dependences $D_{2}(m)$, and the plateaux move down with the growth of the
parameter $Lf_{0}$. As in the experimental case, the plateaux display no
tendency to disappear when $Lf_{0}>100$.

All this enables us to assert that the suggested model has passed the
verification by the fractal analysis procedure, and the model reproduces
qualitatively all important peculiarities. Besides, the model has an
important property: it is not arbitrary in the physical system studied.

\section{Conclusion}

In this paper we have made a constructive attempt to confirm or to reject
various models for describing voltage fluctuations in Sn-Ge thin-film
superconducting composites in the vicinity of the percolation threshold at
temperatures close to the superconducting transition. We did our best to
confine ourselves to studying experimental realizations and to abstain from
speculative theorizing.

We have employed the procedure for fractal analysis. The procedure enables
us, as a minimum, to reject two hypotheses about the nature of the
realizations observed: the model of a determinate dynamic system and the
model of the colored noise with a power-law energy spectrum and random
phases. If one adds to this list obviously inadequate models of the Markov
process, or the process of a small correlation radius, and that of a
determinate system with a simple regular behavior, then one obtains a
complete list of models which can be tested by using constructive procedures.

We believe that our results permit us to reject the colored noise model
reliably. However, the alternative model of a dynamic process is not quite
adequate, since it does not account for the correlated dimension change in
the function of the processing parameters. This demands either a more or
less arbitrary complication of the dynamic model or a complete change of the
viewpoint. We have chosen the latter and, retaining the simple dynamic
nature, suggested another model of the process observed. We believe that our
model successfully interrelates various physical phenomena typical for the
studied system.

We have suggested and verified the model of the voltage fluctuation origin
by a passive transformation of any magnetic field oscillations with the
transformation fractal function. We have studied in detail the fractal
nature of the transition function of such a transforming mechanism for the
Sn-Ge percolation system. Its fractality explains both the amplification of
arbitrarily weak signals to the observable level and a universal spectrum
composition of realizations, even for the simplest detectable processes. We
have shown that the corresponding model realizations manifest all typical
peculiarities of the dependence of the correlated dimension $D_{2}$ on the
dimension of the imbedding and the stationarity parameter $Lf_{0}$.

The fractal character of the transition function seems to be related to the
existence of a wide and self-similar distribution of Josephson contour
areas. Some theoretical models of this phenomenon were reported by Grib \cite
{Grib}. The formation of a transition function having a fractal symmetry
seems to be typical for a broad class of percolation systems (such are, for
example, all the objects at the point of the second order phase transition).
From this point of view the suggested model can be regarded as one of
universal models of generating the ``1/f noise'' in a large series of
physical as also non-physical objects and to confirm the fundamental fractal
nature of the ``1/f noise''. Here we have not discussed the variation
profile of the fractal dimension of the transition function $V_{DC}(H)$ and
the correlated dimensions $D_{2}$ with the variation of temperature. A
separate paper will be devoted to this question. This does not pretend to
exhaust the topic. On the contrary, we intend to start and stimulate a
discussion of the phenomenon.

\section{Acknowledgments}

This work was partially supported by the Fundamental Research Foundation of
the Ukrainian State Committee on Science and Technology.

\newpage

\end{document}